\documentclass[%
reprint,
superscriptaddress,
bibnotes,amssymb,
aps,
floatfix,
longbibliography,
prl]{revtex4-2}
\usepackage{qcircuit}
\usepackage{amsmath,bm}
\usepackage{amsthm}

\makeatletter
\renewenvironment{proof}[1][\proofname]{%
  \par
  \pushQED{\qed}%
  \normalfont
  \topsep6\p@\@plus6\p@\relax
  \trivlist
  \item[\hskip\labelsep\itshape\underline{#1\@addpunct{.}}]\ignorespaces
}{%
  \popQED
  \endtrivlist
  \@endpefalse
}
\makeatother

\makeatletter
\newenvironment{remarks}[1][Remarks]{%
  \par
  \normalfont
  \topsep6\p@\@plus6\p@\relax
  \trivlist
  \item[]%
  \noindent{\itshape\underline{#1\@addpunct{.}}}%
  \par\nobreak\noindent
  \ignorespaces
}{%
  \endtrivlist
  \@endpefalse
}
\makeatother

\usepackage{amsmath,amssymb,mathrsfs,amsfonts,dsfont}
\usepackage{subfigure, epsfig}
\usepackage{braket}
\usepackage{ulem}
\usepackage{enumerate}
\usepackage{color}
\usepackage{graphicx}
\usepackage{appendix}
\usepackage{enumitem}
\usepackage{algorithm}
\usepackage{algpseudocode}
\usepackage{comment}
\usepackage{hyperref}
\newtheorem{theorem}{Theorem}
\newtheorem{corollary}{Corollary}
\newtheorem{lemma}{Lemma}

\newtheorem{example}{Example}

\newcommand{\Tr}{\mathrm{Tr}}

\begin{document}


\title{$N$-Party Hadamard Test for Distributed Quantum Computation}


\author{Kaoru Yamamoto}
\email{kaoru.yamamoto@ntt.com}
\affiliation{NTT Computer and Data Science Laboratories, NTT Corporation, Musashino 180-8585, Japan}

\author{Yuichiro Matsuzaki}
\affiliation{Department of Electrical, Electronic, and Communication Engineering, Faculty of Science and Engineering, Chuo University, 1-13-27 Kasuga, Bunkyo-ku, Tokyo 112-8551, Japan}

\author{Yasunari Suzuki}
\affiliation{NTT Computer and Data Science Laboratories, NTT Corporation, Musashino 180-8585, Japan}

\author{Yuuki Tokunaga}
\affiliation{NTT Computer and Data Science Laboratories, NTT Corporation, Musashino 180-8585, Japan}

\author{Suguru Endo}
\email{suguru.endou@ntt.com}
\affiliation{NTT Computer and Data Science Laboratories, NTT Corporation, Musashino 180-8585, Japan}
\affiliation{JST, PRESTO, 4-1-8 Honcho, Kawaguchi, Saitama, 332-0012, Japan}

\begin{abstract}
Quantum computers promise computational advantages over classical computers, but hardware-imposed limitations remain a major obstacle. 
The Hadamard test mitigates these limitations by estimating expectation values associated with resource-intensive quantum operations using simple quantum circuits at the cost of additional classical sampling, and therefore underlies many quantum algorithms.
However, in distributed quantum computing (DQC), which offers a promising route to scalability, its use is hindered by the need for nonlocal controlled operations. 
Here we introduce an $N$-party Hadamard test for DQC that estimates the same expectation values as the standard Hadamard test without implementing nonlocal controlled operations. The protocol instead uses pre-shared entanglement together with local operations and classical communication, which are standard resources in DQC settings.
To demonstrate its utility, we apply it to unitary operations for clustered Hamiltonian simulation and to projectors for stabilizer-state preparation, showing lower sampling overheads than previous approaches by exploiting pre-shared entangled ancilla states.
Moreover, we numerically demonstrate Bell-state preparation from Werner states to show favorable sampling efficiency and noise robustness relative to conventional purification, circuit knitting/cutting, and probabilistic error cancellation.
Our work provides a general strategy for bringing Hadamard-test-based algorithms to DQC, facilitating practical and flexible quantum computation.
\end{abstract}

\maketitle

\textbf{\textit{Introduction.}--}
Quantum computers promise computational advantages over classical computers, but fault-tolerant quantum computation (FTQC) requires a large number of physical qubits. On the path toward FTQC, the limited quantum resources available on current hardware remain a central obstacle. The Hadamard test is useful in this regime because it can estimate expectation values associated with certain resource-intensive quantum operations using relatively simple quantum circuits, albeit with sampling overhead~\cite{bravyi2016trading}. 
It is therefore used in a broad class of quantum algorithms, particularly near-term and early-FTQC algorithms, including quantum error mitigation~\cite{bonet2018low,endo2020hybrid,cai2022quantum,tsubouchi2023virtual,endo2022quantum,tsubouchi2025symmetric, araki2025correcting} and Hamiltonian and Lindbladian simulation~\cite{yu2021fast,zeng2021universal,wan2022randomized,lin2022heisenberg,jinzhao2022perturbative,dong2023linear,chakraborty2024implementingany,yu2024exponentially,harrow2025optimal,schmitt2025cutting,yu2025lindbladian,zeng2025simple,gunther2025phase,kato2025exponentially,wada2025state,ding2025quantum,sun2025randomised,wada2025tradeoffs,jakob2026phase}.

Distributed quantum computing (DQC) offers a promising route to scaling quantum computers beyond the size of a single processor~\cite{bravyi2022the,ang2024arquin,aghaeeRad2025ModularPhotonicQC}, and its key components have been steadily developed \cite{cabrillo1999creation,bose1999proposal,benjamin2005optical,chou2005measurement,lim2005repeat,barrett2005efficient,benjamin2006brokered,moehring2007entanglement,benjamin2009quantum,sheng2010complete,nickerson2014freely,nigmatullin2016minimally,hu2021long,jnane2022multicore,marcello2024distributed}.
Since practical DQC architectures are also expected to operate with limited quantum resources, the Hadamard test would be highly useful in this setting.
However, a straightforward extension of the Hadamard test to distributed architectures is challenging, because it typically requires nonlocal controlled operations, which are not directly available as native operations in architectures restricted to local operations and classical communication (LOCC).
Such operations can be implemented by using entanglement purification~\cite{bennet1996purification,deutsch1996quantum,dur2007entanglement,fujii2009entanglement} followed by quantum teleportation~\cite{bennett1993teleporting,eisert2000optimal,pirandola2015advances,hu2023progress,wu2023entanglement}.
However, this approach consumes substantial quantum resources and computational time, thereby undermining the resource-saving advantage of the Hadamard test.
Several studies have considered Hadamard tests for DQC~\cite{jinzhao2022perturbative,harrow2025optimal,schmitt2025cutting}, but they do not exploit the pre-shared entanglement available in DQC and therefore incur large sampling overheads.
The absence of a Hadamard-test protocol tailored to DQC hinders the efficient implementation of many useful distributed quantum algorithms.

Here we introduce an $N$-party Hadamard test tailored to DQC that estimates the same expectation values as the standard Hadamard test without using teleportation to implement nonlocal controlled operations (Theorem~1). The protocol instead uses pre-shared entanglement and LOCC, which are standard resources in DQC settings.
We then show that the protocol can estimate expectation values associated with general maps constructed from linear combinations of unitaries (LCU), with a quantifiable sampling overhead (Theorem~2). 
As applications, we study the unitary-LCU case for clustered Hamiltonian simulation (Example~1 and Corollary~1) and the projector-LCU case for stabilizer-state preparation, including Bell-state preparation at the level of expectation values (Example~2 and Corollary~2). These applications achieve lower sampling overheads than previous approaches.
Furthermore, our numerical demonstration of Bell-state preparation in expectation values shows advantages over representative alternatives: higher fidelity than the conventional double-selection protocol \cite{fujii2009entanglement}, lower sampling overhead than optimal circuit knitting/cutting \cite{piveteau2022circuit}, and greater robustness to fluctuations in the infidelity of shared noisy Bell states than probabilistic error cancellation (PEC) \cite{temme2017error,endo2018practical,yuan2023virtual,takagi2024virtual}. Overall, our work provides a general strategy for bringing Hadamard-test-based algorithms to DQC and serves as a building block for distributed quantum algorithms.

\textbf{\textit{$N$-party Hadamard test.}--}
\begin{figure*}[htbp]
    \centering
    \includegraphics[width= 0.8\linewidth]{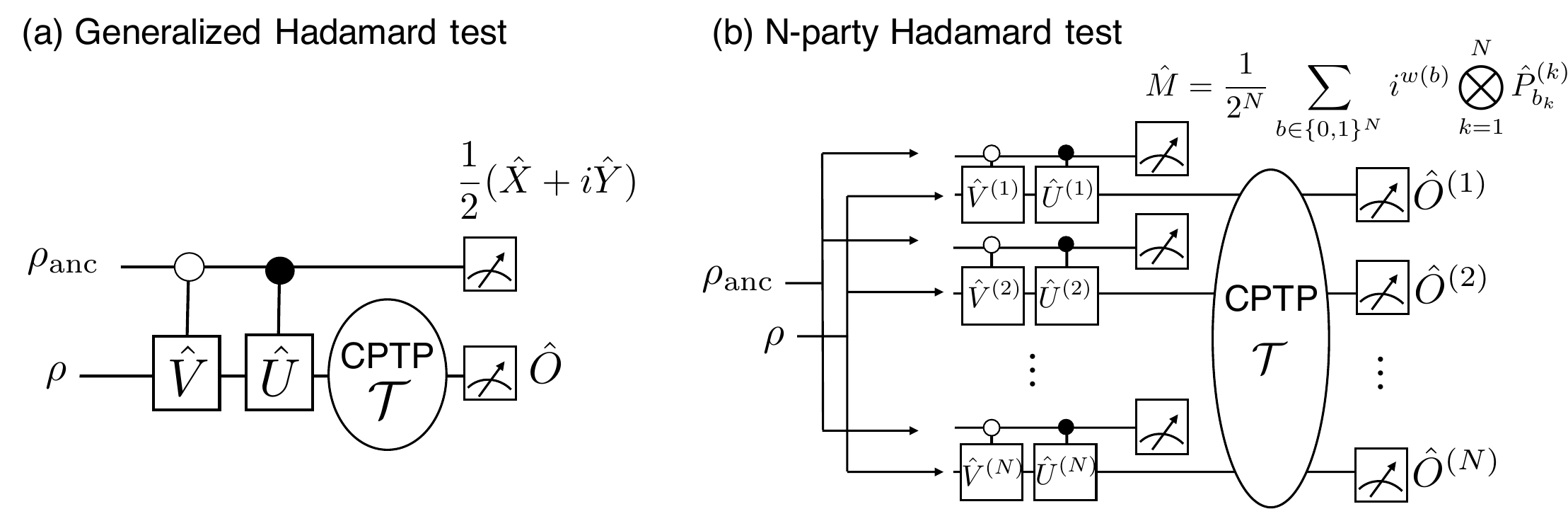}
    \caption{(a) A quantum circuit for the standard generalized Hadamard test. The joint measurement estimates the expectation value associated with the generalized map
    $\rho \mapsto \mathcal{T}(\hat{U}\rho\hat{V}^\dagger)$ as
    $\braket{(\hat{X}+i\hat{Y})/2\otimes\hat{O}}= \rho_\text{anc}^{10}\Tr[\hat{O}\mathcal{T}(\hat{U}\rho\hat{V}^\dagger)]$, where $\rho_\text{anc}^{10}=\braket{1|\rho_\text{anc}|0}$ and $\mathcal{T}$ is a CPTP map.
    (b) A quantum circuit for the $N$-party Hadamard test. The expectation value of the joint measurement estimates $\braket{\hat{M}\otimes\hat{O}}=c\Tr[\hat{O}\mathcal{T}(\hat{U}\rho\hat{V}^\dagger)]$, where $c=\braket{1^{\otimes N}|\rho_\text{anc}|0^{\otimes N}}$, $\hat{O} = \bigotimes_{k=1}^N \hat{O}^{(k)}$, $\hat{U} = \bigotimes_{k=1}^N \hat{U}^{(k)}$, $\hat{V} = \bigotimes_{k=1}^N \hat{V}^{(k)}$, and $\mathcal{T}$ is a CPTP map.
    In both circuits, the black and white circles indicate controls that acts on the target qubit when the ancilla qubit is in the state $\ket{1}$ and $\ket{0}$, respectively.
    }
   \label{fig:1}
\end{figure*}
Before introducing the $N$-party Hadamard test, we briefly review the standard Hadamard test \cite{jinzhao2022perturbative, sun2025randomised, kato2025exponentially, yu2025lindbladian,wada2025state} with an arbitrary single-qubit ancilla.
Figure~\ref{fig:1}(a) shows the corresponding circuit, where $\mathcal{T}$ is an arbitrary CPTP map followed by the measurement of a system observable $\hat{O}$.
The joint expectation value satisfies $\braket{(\hat{X}+i\hat{Y})/2\otimes\hat{O}}
= \rho_\text{anc}^{10}\Tr[\hat{O}\mathcal{T}(\hat{U}\rho\hat{V}^\dagger)]$, where $\rho_\text{anc}^{10}=\braket{1|\rho_\text{anc}|0}$.
The non-Hermitian observable $(\hat{X}+i\hat{Y})/2$ is estimated by Monte Carlo sampling of $\hat{X}$ and $\hat{Y}$ measurements.

We now adapt this idea to DQC, where pre-shared entanglement is allowed but all subsequent operations are restricted to LOCC.
\begin{theorem}
($N$-party Hadamard test)
Let $N$ parties share an ancilla state $\rho_\text{anc}$ with
$c=\braket{1^{\otimes N}|\rho_\text{anc}|0^{\otimes N}}$, together with a distributed input state $\rho$.
The circuit in Fig.~\ref{fig:1}(b) estimates $c\mathrm{Tr}[\hat{O}\mathcal{T}(\hat{U}\rho\hat{V}^\dagger)]$ using only LOCC.
Here, $\hat{U}=\bigotimes_{k=1}^N \hat{U}^{(k)}$ and $\hat{V}=\bigotimes_{k=1}^N \hat{V}^{(k)}$ are products of local unitaries, $\hat{O}=\bigotimes_{k=1}^{N}\hat{O}^{(k)}$ is a product observable, and $\mathcal{T}$ is the CPTP map induced by the subsequent deterministic LOCC process.
\end{theorem}
\begin{proof}
Immediately before the ancilla measurements in Fig.~\ref{fig:1}(b), the relevant off-diagonal component of the state is $\rho_\mathrm{bfm}=c\ket{1^{\otimes N}}\bra{0^{\otimes N}}\otimes\mathcal{T}(\hat{U}\rho\hat{V}^{\dagger})+\cdots$.
To extract the desired value, we consider the operator $\hat{M}=\ket{0^{\otimes N}}\bra{1^{\otimes N}}$.
Then $\braket{\hat{M}\otimes \hat{O}}=\mathrm{Tr}[(\hat{M}\otimes\hat{O})\rho_\mathrm{bfm}]=c\,\mathrm{Tr}[\hat{O}\mathcal{T}(\hat{U}\rho\hat{V}^\dagger)]$.
Although $\hat{M}$ is not itself a directly measurable Hermitian observable, it can be decomposed into a linear combination of local Pauli measurements as
\begin{equation}
\hat{M}
=\ket{0^{\otimes N}}\bra{1^{\otimes N}}  = 
\frac{1}{2^N}
\sum_{b\in\{0,1\}^N}
i^{w(b)}
\bigotimes_{k=1}^N \hat{P}_{b_k}^{(k)},
\label{eq:M}
\end{equation}
where $w(b)=\sum_{k=1}^N b_k$, $\hat{P}_0^{(k)}=\hat{X}^{(k)}$, and $\hat{P}_1^{(k)}=\hat{Y}^{(k)}$.
This follows from expanding $\hat{M} = \ket{0^{\otimes N}}\bra{1^{\otimes N}}=2^{-N}\bigotimes_{k=1}^N(\hat{X}^{(k)}+i\hat{Y}^{(k)})$.
Therefore, $\braket{\hat{M}\otimes\hat{O}}$ can be estimated by Monte Carlo sampling over the local Pauli measurement settings in Eq.~\eqref{eq:M}, which yields the desired value using only LOCC.
\end{proof}
\begin{remarks}
\begin{itemize}[
  labelindent=0pt,
  leftmargin=*,
  labelsep=.5em,
  topsep=0pt,
  partopsep=0pt,
  parsep=0pt,
  itemsep=0.05cm
]
    \item  We can estimate the real and imaginary parts by $\braket{\hat{M}_\text{Re}\otimes\hat{O}}$ and $\braket{\hat{M}_\text{Im}\otimes\hat{O}}$, where
    $\hat{M}_{\rm Re}=(\hat{M}+\hat{M}^\dagger)/2=2^{-N}\sum_{w(b)\,\mathrm{even}}(-1)^{w(b)/2}\!\bigotimes_{k=1}^N\hat P_{b_k}^{(k)}$ and
    $\hat M_{\rm Im}=(\hat{M}-\hat{M}^\dagger)/(2i)=2^{-N}\sum_{w(b)\,\mathrm{odd}}(-1)^{(w(b)-1)/2}\!\bigotimes_{k=1}^N\hat P_{b_k}^{(k)}$ \cite{M_EG}.
    \item When only one of the real and imaginary parts is needed, we can implement $2\hat{M}_{\mathrm{Re}}$ or $2\hat{M}_{\mathrm{Im}}$ by Monte Carlo sampling and thereby estimate $2\mathrm{Re}\big(c\,\Tr[\hat O\,\mathcal T(\hat U\rho\hat V^\dagger)]\big)$ or $2\mathrm{Im}\big(c\,\Tr[\hat O\,\mathcal T(\hat U\rho\hat V^\dagger)]\big)$, which will be used in Theorem~2.
    \item Taking the input to be $\sigma^{\otimes n}$, $\hat{U}$ to be the derangement operator \cite{koczor2021exponential}, and $\hat{V}=\hat{I}$ leads to the generalized SWAP test \cite{knorzer2025distributed} among $N$ parties.
\end{itemize}
\end{remarks}

\textbf{\textit{General maps in expectation values.}--}
Given a linear-combination-of-unitaries (LCU) decomposition of an operator
$\hat{A}$ into product unitaries, the $N$-party Hadamard test realizes the normalized general map
$\rho \mapsto \hat{A}\rho\hat{A}^\dagger/
\mathrm{Tr}[\hat{A}\rho\hat{A}^\dagger]$
in expectation values, even under the DQC restriction, with a quantifiable sampling overhead.
\begin{theorem}(LCU-based general map among $N$ parties)
Let
$\hat{A}=\sum_i a_i\hat{U}_i
=\sum_i a_i\bigotimes_{k=1}^N\hat{U}_i^{(k)}$
be a bounded operator with $a_i\in\mathbb{R}$ \cite{real_ai}, where
$\hat{U}_i^{(k)}$ is a local unitary on the $k$th party.
Suppose that the parties share $\rho_\mathrm{anc}$ before the computation,
that at least one of $2\mathrm{Re}(c)$ and $2\mathrm{Im}(c)$ is nonzero,
and that $\mathrm{Tr}[\hat{A}\rho\hat{A}^\dagger]>0$.
Then, for any nonzero component $C\in\{2\mathrm{Re}(c),2\mathrm{Im}(c)\}$ and for any product observable $\hat{O}=\bigotimes_{k=1}^N\hat{O}^{(k)}$ with $\|\hat{O}\|= 1$ \cite{Onorm}, there exists an LOCC protocol, based on the $N$-party Hadamard test and classical post-processing, that estimates
\begin{equation}
\frac{\mathrm{Tr}[\hat{O}\mathcal{T}(\hat{A}\rho\hat{A}^\dagger)]}
{\mathrm{Tr}[\hat{A}\rho\hat{A}^\dagger]}.
\label{eq:genrho}
\end{equation}
The sampling overhead of this estimator is $\mathcal{O}(\gamma^2)$, where
$\gamma=(\mathrm{Tr}[\hat{A}\rho\hat{A}^\dagger])^{-1}|C|^{-1}\|\mathbf{a}\|_1^2$
with $\|\mathbf{a}\|_p=(\sum_i |a_i|^p)^{1/p}$.
\end{theorem}
\begin{proof}
We provide an explicit estimator for Eq.~\eqref{eq:genrho}.
\begin{enumerate}[
  labelindent=0pt,
  leftmargin=*,
  labelsep=.5em,
  topsep=0pt,
  partopsep=0pt,
  parsep=0pt,
  itemsep=0.05cm
]
    \item Sample $i,j$ with nonzero $a_i a_j$ according to
    $p_{ij}=|a_i a_j|/\|\mathbf{a}\|_1^2$ and run the $N$-party Hadamard test with
    $\hat{U}=\hat{U}_i=\bigotimes_{k=1}^N\hat{U}_i^{(k)}$ and
    $\hat{V}=\hat{U}_j=\bigotimes_{k=1}^N\hat{U}_j^{(k)}$.
    Using either $2\hat{M}_\mathrm{Re}$ or $2\hat{M}_\mathrm{Im}$, we obtain the corresponding real- or imaginary-part estimator, $2\mathrm{Re}\big(...\big)$ or $2\mathrm{Im}\big(...\big)$, respectively.
    \item Multiply this value by
    $\|\mathbf{a}\|_1^2 a_i a_j/|a_i a_j|$ and average over $i,j$.
    This gives an estimator of the numerator, $\sum_{i,j}
    \frac{\|\mathbf{a}\|_1^2 a_i a_j}{|a_i a_j|}
    p_{ij}
    \braket{2\hat{M}_\mathrm{Re(Im)}\otimes\hat{O}}_{ij}
    =
    C\,\mathrm{Tr}[\hat{O}\mathcal{T}(\hat{A}\rho\hat{A}^\dagger)]$ \cite{supple}.
    The multiplication increases the variance by a factor of $\|\mathbf{a}\|_1^4$.
    \item Estimate the denominator from the same measurement data by discarding the measurement outcome of $\hat{O}$, which is equivalent to setting $\hat{O}=\hat{I}$.
    Since $\mathcal{T}$ is trace-preserving, this gives
    $C\,\mathrm{Tr}[\mathcal{T}(\hat{A}\rho\hat{A}^\dagger)]
    =
    C\,\mathrm{Tr}[\hat{A}\rho\hat{A}^\dagger]$.
    No additional quantum processing is required.
    \item The ratio of the numerator and denominator estimators gives Eq.~\eqref{eq:genrho}.
    \end{enumerate}
In the large-sample limit, the variance of the ratio estimator $\bar{x}/\bar{y}$, constructed from the sample means $\bar{x}$ and $\bar{y}$, is given by $\mathrm{Var}(\bar{x}/\bar{y}) = \sigma_\mathrm{ratio}^2N_{\mathrm{samp}}^{-1} + \mathcal{O}(N_{\mathrm{samp}}^{-2})$, where $\sigma_\mathrm{ratio}^2 = \mu_y^{-2}[\mathrm{Var}(x) + \mu_x^2\mu_y^{-2}\mathrm{Var}(y) - 2\mu_x\mu_y^{-1}\mathrm{Cov}(x,y)]$, $N_{\mathrm{samp}}$ is the number of samples, $\mu_x$ and $\mu_y$ denote the means of $x$ and $y$, respectively, and $\mathrm{Cov}(x,y)$ denotes their covariance \cite{vanderVaart1998asymptotic}. Therefore, relative to a standard quantum observable, the required number of samples is increased by a factor determined by $\sigma_\mathrm{ratio}^2$.
For the protocol described above, this expression gives $\sigma_\mathrm{ratio}^2= \mathcal{O}(\gamma^2)$ up to constant factors. This scaling follows because $\mathrm{Var}(x)$ and $\mathrm{Var}(y)$ scale as $\|\mathbf{a}\|_1^4$ due to the multiplication performed in Step 2, and $\mu_y = C\,\mathrm{Tr}[\hat{A}\rho\hat{A}^{\dagger}]$.
\end{proof}
\begin{remarks}
\begin{itemize}[
  labelindent=0pt,
  leftmargin=*,
  labelsep=.5em,
  topsep=0pt,
  partopsep=0pt,
  parsep=0pt,
  itemsep=0.05cm
]
    \item The maximum value $|C|=1$ follows from the non-negativity of the density matrix and is achieved, for example, by an $N$-qubit GHZ ancilla state \cite{jinmin2023unified, araki2025space}.
    \item The protocol does not require prior knowledge of $\rho_\text{anc}$ or $C$, because the same factor $C$ appears in both the numerator and denominator estimators. If $C$ is known beforehand, the sampling overhead can be reduced \cite{supple}.
    \item Using the $N$-party generalized SWAP test described in the remarks following Theorem~1, one can implement SWAP-test-based algorithms such as virtual channel purification \cite{liu2025virtual}.
\end{itemize}
\end{remarks}

To show the utility of the $N$-party Hadamard test, we give two examples below, corresponding to the cases where
$\hat{A}$ is a unitary operator and a projector.
The corresponding corollaries show that the sampling overhead can be lower than in previous approaches.
\begin{example}(Unitary LCU case)
When the operator $\hat{A}=\sum_i a_i\hat{U}_i$ in Theorem~2 is unitary,
the protocol of Theorem~2 can estimate expectation values after the unitary
evolution $\rho\mapsto \hat{A}\rho\hat{A}^\dagger$ among $N$ parties. Since
$\mathrm{Tr}[\hat{A}\rho \hat{A}^\dagger]=1$, the sampling-cost factor is $\gamma=|C|^{-1}\|\mathbf{a}\|_1^2$.
\end{example}

In related unitary-LCU settings, several previous studies have analyzed
Hadamard-test-type primitives with a $\ket{+}$ ancilla, focusing mainly on
the sampling overhead associated with the LCU one-norm
$\|\mathbf{a}\|_1$ \cite{jinzhao2022perturbative,harrow2025optimal,schmitt2025cutting}.
For an $N$-party DQC extension, however, another source of overhead becomes important: the ancilla-state-dependent factor $C$. 
Indeed, restricting the ancilla resource to separable states can make
$|C|^{-2}$ grow exponentially with $N$. 
For a separable ancilla state such as
$\rho_\text{anc} = (\ket{+}\bra{+})^{\otimes N}$, the sampling overhead
increases exponentially with $N$ because $\gamma=|C|^{-1}\|\mathbf{a}\|_1^2 =2^{(N-1)}\|\mathbf{a}\|_1^2$, where we used $C = 2\mathrm{Re}(c) = 2\braket{1^{\otimes N}|\rho_\text{anc}|0^{\otimes N}} =2^{-(N-1)}$.
Our formulation with entangled ancilla states can mitigate this unfavorable
$N$-dependence; for example, an ideal $N$-qubit GHZ ancilla gives $C=1$, independent of $N$.
The following corollary for clustered Hamiltonian simulation, a setting studied
in circuit cutting/knitting
\cite{peng2020simulating,harrow2025optimal}
and perturbative quantum simulation
\cite{jinzhao2022perturbative}, makes these points explicit. 
\begin{corollary}
($N$-party clustered Hamiltonian simulation)
Suppose that the Hamiltonian is decomposed as
$\hat{H}=\hat{H}_\mathrm{loc}+\sum_{j\in\partial A}\hat{H}_j$
with respect to a partition of the system into $N$ disjoint subsets
$A_1,A_2,\ldots,A_N$, where $\partial A$ denotes the set of boundary terms coupling different subsets.
Here, $\hat{H}_\mathrm{loc}=\sum_{k=1}^N\hat{H}^{(k)}$ acts within the individual subsets.
Assume that the boundary terms satisfy $\|\hat{H}_j\|\leq 1$ and that
$\hat{H}_j/\|\hat{H}_j\|$ is a product unitary across the subsets involved in the interaction, with identities acting on subsets not involved in the interaction. Then, for any product observable across the partition with $\|\hat{O}\|= 1$ \cite{Onorm}, the expectation value $\mathrm{Tr}[\hat{O}e^{-i\hat{H}t}\rho e^{i\hat{H}t}]$ can be estimated using the $N$-party Hadamard test with the quantum circuit shown in Fig.~\ref{fig:QCinEM}(a). The sampling overhead is at most $\mathcal{O}(|C|^{-2}e^{4\eta t})$, where $\eta=\sum_{j\in\partial A}\|\hat{H}_j\|$.
\end{corollary}
\begin{proof}
The proof is given in the End Matter.
\end{proof}

The second example is obtained by taking $\hat{A}$ to be a projector. 
In particular, taking $\hat{A}$ to be a stabilizer projector is useful because such a projector has an LCU decomposition with $\|\mathbf{a}\|_1^2=1$ as shown below, and hence the sampling overhead is not amplified by a large LCU one-norm.
\begin{example}
(Stabilizer-projector LCU case)
Let $\hat{G}_i \in \mathbb{G}$ be a stabilizer generator and $\hat{S}_i\in\mathbb{S}$ be a stabilizer operator, where $|\mathbb{S}|$ is the number of stabilizer operators.
Taking $\hat{A}=\hat{P}$ to be the stabilizer projector, we can decompose it as a linear combination of stabilizer operators as
$\hat{A} = \prod_{\hat{G}_i \in \mathbb{G}} (\hat{I}+\hat{G}_i)/2 = |\mathbb{S}|^{-1}\sum_{\hat{S}_i\in\mathbb{S}} \hat{S}_i$ \cite{macclean2020decoding,tsubouchi2023virtual,endo2022quantum}.
Since each stabilizer is a product of local unitaries, Theorem~2 applies. Because $\|\mathbf{a}\|_1^2 = 1$, the sampling overhead is $\gamma^2 = |C|^{-2}F^{-2}$, where $F = \mathrm{Tr}[\hat{P}\rho\hat{P}]$ is the projection probability onto the stabilizer subspace.
\end{example}
We remark that for a non-stabilizer target state that admits a stabilizer decomposition, one can evaluate each stabilizer state using Example~2 and classically combine the results with an additional sampling overhead of $R^2$, where $R$ is the robustness of magic associated with the decomposition \cite{howard2017application}. Importantly, this decomposition is applied to the target state after projection, rather than to the projector itself.

We use Example~2 to prepare entangled stabilizer states in expectation values.
In particular, Bell-state preparation is relevant to practical DQC because Bell states are basic resources for implementing nonlocal operations via teleportation.
Although our protocol does not physically output purified Bell states, it provides the corresponding expectation values whenever the final goal is expectation-value estimation after a deterministic distributed protocol.
Indeed, any deterministic use of Bell states, including deterministic teleportation with feed-forward corrections, is described by a CPTP map and can therefore be absorbed into $\mathcal{T}$ in Eq.~\eqref{eq:genrho}.
Thus, our protocol reproduces the expectation values that would be obtained after consuming purified Bell states, at the cost of sampling overhead rather than physical purification.

Preparing $n$ Bell states at the level of expectation values for DQC via quasi-probability decomposition (QPD) has been studied in the contexts of circuit knitting/cutting \cite{piveteau2022circuit,vazquez2024scaling} and virtual resource distillation \cite{yuan2023virtual,takagi2024virtual,zhang2024experimental}. 
However, for QPD-based methods, finding an explicit and implementable quantum circuit that satisfies the LOCC restriction while keeping the sampling overhead small becomes increasingly difficult as $n$ grows.
In contrast, our protocol provides an explicit quantum circuit for arbitrary $n$ with a sampling overhead that matches the optimal QPD-based overhead up to constant factors \cite{piveteau2022circuit,yuan2023virtual,takagi2024virtual}, as shown in the following corollary and remarks. Thus, our protocol with the $N$-party Hadamard test provides an efficient and practical way to prepare $n$ Bell states at the level of expectation values.
\begin{corollary}
(Bell-state preparation in expectation values)
Taking $\mathbb{S}=\{\hat{I}\otimes\hat{I},\hat{X}\otimes\hat{X},
-\hat{Y}\otimes\hat{Y},\hat{Z}\otimes\hat{Z}\}^{\otimes n}$,
Example~2 prepares $n$ Bell states, $\rho_\text{Bell}^{\otimes n} = (\ket{\Phi_\text{Bell}}\bra{\Phi_\text{Bell}})^{\otimes n}$ with $\ket{\Phi_\text{Bell}} = (\ket{00}+\ket{11})/\sqrt{2}$, in expectation values
with $\gamma=|C|^{-1}F^{-1}$.
\end{corollary}
\begin{proof}
The stabilizer group of $n$ Bell states is
$\mathbb{S}=\{\hat{I}\otimes\hat{I},\hat{X}\otimes\hat{X},
-\hat{Y}\otimes\hat{Y},\hat{Z}\otimes\hat{Z}\}^{\otimes n}$.
Therefore, the claim follows directly from Example~2.
\end{proof}

\begin{remarks}[Remarks]
\begin{itemize}[
  labelindent=0pt,
  leftmargin=*,
  labelsep=.5em,
  topsep=0pt,
  partopsep=0pt,
  parsep=0pt,
  itemsep=0.05cm
]
\item For separable ancilla and input states, the maximum coherence factor is
$|C|=1/2$, achieved for example by
$\rho_\mathrm{anc}=(\ket{+}\bra{+})^{\otimes 2}$, and the Bell-state fidelity satisfies $F\leq 2^{-n}$, with equality achieved for example by $(\ket{+}\bra{+})^{\otimes 2n}$.
Thus, the best separable-input and separable-ancilla case gives $\gamma=2^{n+1}$, comparable to the optimal circuit-knitting/cutting cost based on QPD, $\gamma=2^{n+1}-1$ \cite{piveteau2022circuit}.
\item For entangled ancillae and input states, the sampling-cost factor of our protocol is comparable to, and in some regimes may even be smaller than, the optimal QPD-based cost $\gamma=2F^{-1}-1$ \cite{yuan2023virtual,takagi2024virtual}. This does not lead to a contradiction because our protocol is not based on QPD. It would therefore be an interesting direction for future work to analyze the optimal sampling overhead of virtual resource distillation based on the $N$-party Hadamard test from a resource-theoretic perspective.
\end{itemize}
\end{remarks}

\textbf{\textit{Numerical demonstration.}--}
We numerically demonstrate Bell-state preparation in expectation values using Qulacs \cite{suzuki2021qulacs}. 
The goal is to illustrate three practical features of our protocol: high output fidelity compared with conventional entanglement purification, reduced sampling overhead compared with circuit knitting/cutting, and robustness to imperfect noise knowledge compared with probabilistic error cancellation (PEC). 
We consider preparing $n$ Bell states from $n+1$ Werner states, using $\rho=\rho_\text{Werner}^{\otimes n}$ as the input and $\rho_\text{anc}=\rho_\text{Werner}$ as the shared ancilla, where $\rho_\text{Werner}=(1-4\epsilon/3)\rho_\text{Bell}+\epsilon\hat{I}/3$ \cite{Werner}. Details of the local noise setting are provided in the End Matter.
\begin{figure*}
    \centering
    \includegraphics[width=\linewidth]{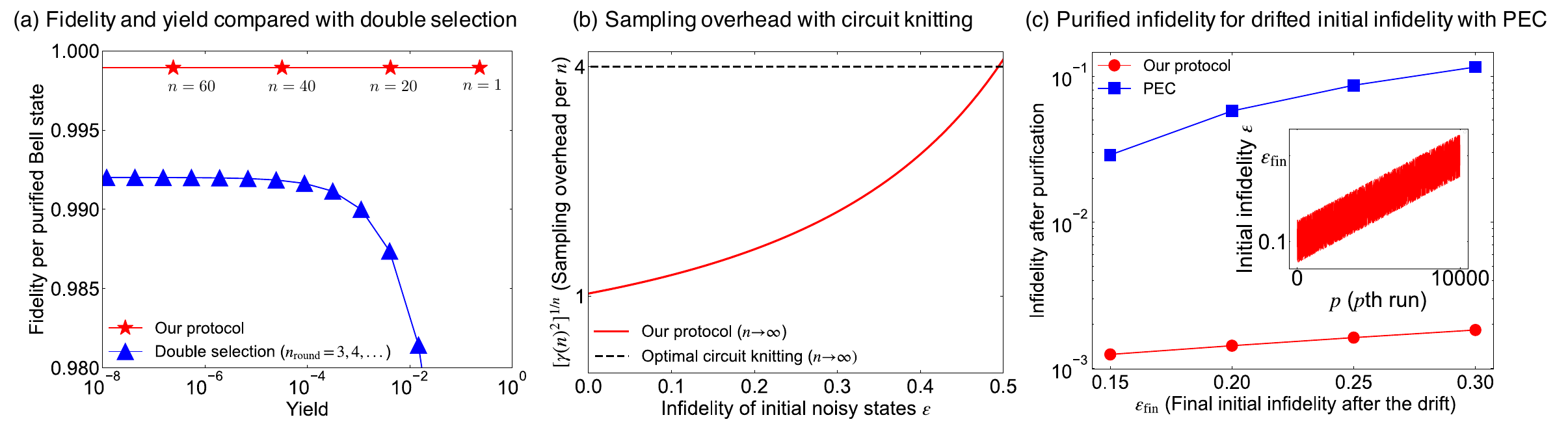}
    \caption{(a) Fidelity per purified Bell state with the corresponding effective yield for our protocol (red stars) and the conventional double-selection protocol (blue curve with triangles). 
    (b) Sampling overhead per purified Bell state, $[\gamma(n)^2]^{1/n}$, as a function of the initial infidelity $\epsilon$ of the input Werner states.  
    The black dotted horizontal line represents the asymptotic optimal overhead of circuit knitting/cutting, $\lim_{n\to\infty}[(2^{n+1}-1)^2]^{1/n}=4$ \cite{piveteau2022circuit,brenner2023optimal,harada2023optimal}.
    (c) Infidelity after purification for our protocol and PEC. The noisy input state is averaged over $M$ runs as described in the main text, with the infidelity drift shown in the inset.}
   \label{fig:numerics_ED}
\end{figure*}

Figure~\ref{fig:numerics_ED}(a) compares the fidelity and effective yield of our protocol with those of the conventional double-selection protocol \cite{fujii2009entanglement}; see the End Matter for numerical details. 
Our protocol achieves a fidelity of 99.9\%, higher than the maximum fidelity of 99.2\% achieved by the double-selection protocol, while retaining a comparable effective yield for several dozen Bell states.
This higher fidelity arises from classical post-processing: averaging and normalization suppress local two-qubit-gate errors that do not affect the ancillae, such as $\hat{I}\otimes\hat{P}$ with $\hat{P}=\hat{X},\hat{Y},\hat{Z}$.
Such errors cannot be detected by double selection, which relies on one-shot error-detection measurements on the ancillae \cite{fujii2009entanglement,krastanov2019optimized}.
A similar advantage is observed for other local noise models, such as dephasing and amplitude-damping noise, as shown in Fig.~\ref{fig:numerics_noise} in the End Matter.

Figure~\ref{fig:numerics_ED}(b) compares the sampling overhead per Bell state, $[\gamma(n)^2]^{1/n}$, with the optimal overhead of circuit knitting/cutting, a standard approach for practical DQC \cite{vazquez2024combining,gupta2025gate}. 
Given access to noisy shared Bell states, our protocol substantially reduces the sampling overhead for reasonable initial infidelities.
For example, at $\epsilon=0.1$, which may be achievable with current technologies \cite{chou2018deterministic,campagne2018deterministic,kannan2020generating,daiss2021a,leent2022entangling,luo2022postselected,leung2022deterministic,yan2022entanglement,chan2023on-chip,qiu2023deterministic}, our protocol gives $[\gamma(n)^2]^{1/n}=1.28$, compared with the optimal circuit-knitting/cutting value $[\gamma(n)^2]^{1/n}=4$ \cite{piveteau2022circuit,brenner2023optimal,harada2023optimal}.
This corresponds to reducing the total sampling overhead to approximately $(1.28/4)^n\simeq(1/3)^n$ of the circuit-knitting/cutting overhead for simulating $n$ purified Bell states.

Moreover, as shown below, our protocol is error-agnostic and robust against imperfect noise knowledge.
This follows from the fact that our protocol does not require knowledge of the ancilla state, together with the following observation.
Let $\rho_p$ denote the input state in the $p$th experimental run, and let $M$ be the total number of runs.
Assuming independent fluctuations in the input state, the effective noisy input state is described by the averaged state
$\bar{\rho}=M^{-1}\sum_{p=1}^{M}\rho_p$.
In the presence of run-to-run drift and fluctuations, the expectation value obtained by our protocol is
\begin{equation}
\frac{
\mathrm{Tr}\!\left[
\hat{O}\mathcal{T}
(\hat{P}\bar{\rho}\hat{P})
\right]
}{
\mathrm{Tr}\!\left[
\hat{P}\bar{\rho}
\hat{P}\right]}.
\end{equation}
Thus, our protocol effectively purifies the averaged noisy resource states and does not require estimating the individual infidelity in each run.

This noise-agnostic feature is important in experiments, where the quality of shared Bell states may drift or fluctuate over time.
Noise-inversion-based QEM methods, such as PEC, require a calibrated noise parameter to construct the inverse map; if the actual noise changes after calibration, the inverse map no longer matches the physical noise and causes a systematic bias.
By contrast, our protocol requires no prior knowledge of the resource-state infidelity, because it virtually applies the stabilizer projector directly to the noisy resource states sampled during the experiment.

To illustrate this robustness, Fig.~\ref{fig:numerics_ED}(c) compares our protocol with PEC for $n=1$; see the End Matter for details of PEC.
We fix the infidelity used to construct the PEC inverse map to the pre-calibrated value $\epsilon=0.1$, while the actual infidelity drifts from $0.1$ to $\epsilon_{\rm fin}$ with an additional uniformly distributed random fluctuation drawn from $[-0.05,0.05]$, as shown in the inset.
Using $M=10^4$ samples, our protocol achieves a lower purified infidelity than PEC when the actual infidelity deviates from the pre-calibrated value.
This result highlights the practical advantage of the error-agnostic stabilizer-projection mechanism in the presence of unknown drift and fluctuations in shared noisy Bell states.

\textbf{\textit{Conclusion and outlook.}--} 
In conclusion, we proposed an $N$-party Hadamard test for DQC using pre-shared entanglement and LOCC, and showed that it realizes general maps in expectation values with a quantifiable sampling overhead.
To show its utility, we applied this framework to the unitary-LCU case for clustered Hamiltonian simulation and the projector-LCU case for entangled stabilizer-state preparation, showing lower sampling
overhead than previous approaches.
Numerical simulations of Bell-state preparation demonstrate favorable sampling efficiency and robustness relative to entanglement purification, circuit knitting/cutting, and PEC.
These results provide a general route to bringing Hadamard-test-based algorithms to distributed architectures.
An important future direction is to clarify the resource tradeoffs enabled by entangled assistance, including those among pre-shared entanglement, sampling overhead, the number of parties, and the locality of the target operations.

\begin{acknowledgments}
\textit{Acknowledgments.---}
This work was supported by JST [Moonshot R\&D] Grant Nos.~JPMJMS2061 and JPMJMS226C; JST, PRESTO, Grant No.~JPMJPR2114, Japan; MEXT Q-LEAP, Grant Nos.~JPMXS0120319794 and JPMXS0118068682; JST CREST Grant No.~JPMJCR23I4; JSPS KAKENHI, Grant No.~23H04390.
Y.~M.~is supported by JSPS KAKENHI (Grant No.~23H04390) and PRESTO, JST. This work was also supported by CREST
(JPMJCR23I5), JST.
\end{acknowledgments}
\bibliography{bib}

\begin{thebibliography}{94}
\bibitem [{rea()}]{real_ai}%
  \BibitemOpen
  \href@noop {} {}\bibinfo {note} {Complex coefficients can be included by
  absorbing their phases into the corresponding unitaries, and hence we take
  $a_i \in \mathbb{R}$ without loss of generality.}\BibitemShut {Stop}%
\bibitem [{Ono()}]{Onorm}%
  \BibitemOpen
  \href@noop {} {}\bibinfo {note} {When $\|O\| \neq 1$, we can easiliy show
  that
  $\gamma=(\mathrm{Tr}[\hat{A}\rho\hat{A}^\dagger])^{-1}|C|^{-1}\|\mathbf{a}\|_1^2\|\hat{O}\|$.}\BibitemShut
  {Stop}%
\end{thebibliography}

\clearpage
\section*{End Matter} 
\appendix

\section{Proof of Corollary~1}
\begin{figure}
    \centering
    \includegraphics[width=0.85\linewidth]{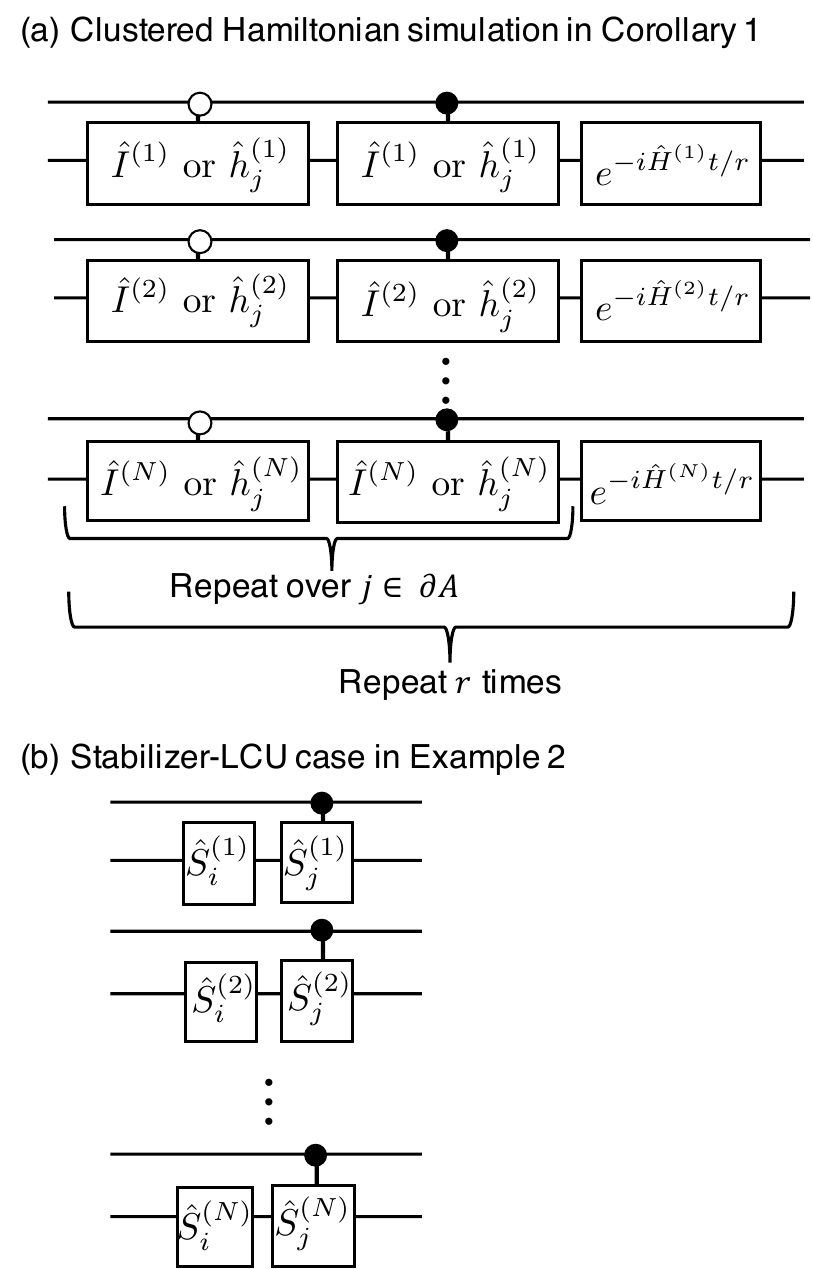}
    \caption{Controlled-operation blocks in the quantum circuit for the $N$-party Hadamard test shown in Fig.~\ref{fig:1}(b): 
(a) the clustered Hamiltonian simulation of Corollary~1 and 
(b) the stabilizer-state preparation of Example~2, where the controlled operations are reduced using the closure of the stabilizer group, $\hat{S}_{i}^{}\hat{S}_j=\hat{S}_k\in\mathbb{S}$ \cite{tsubouchi2023virtual}.}
   \label{fig:QCinEM}
\end{figure}

\begin{proof}
To apply Theorem~2, we first construct a product-unitary LCU for the Trotterized time evolution, 
$\hat{U}_r(t)=(e^{-i\hat{H}_\mathrm{loc}t/r}
\prod_{j\in\partial A}e^{-i\hat{H}_j t/r})^r$ for a positive integer $r$, so that
$\hat{U}_r(t)\to e^{-i\hat{H}t}$ as $r\to\infty$.
Using $e^{-i\hat{H}_j t/r}=\cos(x_j)\hat{I}
+\sin(x_j)(-i\hat{H}_j/\|\hat{H}_j\|)$ with $x_j \equiv \|\hat{H}_j\|t/r$ for each boundary term,
we obtain
\begin{align}
&\hat{U}_r(t)=(e^{-i\hat{H}_\mathrm{loc}t/r}\prod_{j\in\partial A}e^{-i\hat{H}_j t/r})^r \notag \\
&= \left[e^{-i\hat{H}_\mathrm{loc}t/r}\prod_{j\in\partial A}\left[\cos(x_j)\hat{I}+\sin(x_j)(-i\hat{H}_j/\|\hat{H}_j\|)\right]\right]^r \notag \\
&\equiv \sum_i a_{i,r}\bigotimes_{k=1}^{N}\hat{U}_{i,r}^{(k)} \label{eq:EMcol1}
\end{align}
The third line follows because each factor
$e^{-i\hat{H}_jt/r}=\cos(x_j)\hat{I}
+\sin(x_j)(-i\hat{H}_j/\|\hat{H}_j\|)$
is an LCU of local product unitaries, multiplying these decompositions, together with $e^{-i\hat{H}_\mathrm{loc}t/r}
=\bigotimes_{k=1}^N e^{-i\hat{H}^{(k)}t/r}$,
gives an LCU of product unitaries $\hat{U}_r(t)\equiv \sum_i a_{i,r}\bigotimes_{k=1}^{N}\hat{U}_{i,r}^{(k)}$. 
Each product of local unitaries $\bigotimes_{k=1}^{N}\hat{U}_{i,r}^{(k)}$ can be implemented by the quantum circuit shown in Fig.~\ref{fig:QCinEM}, where the global phase $-i$ can be absorbed into one of the local unitary factors.
Then the protocol of Theorem~2 in the unitary LCU case can estimate $\mathrm{Tr}[\hat{O}\hat{U}_r(t)\rho\hat{U}_r^\dagger(t)]$.

The next step is to bound the coefficient one-norm of this expansion.
Note that $a_{i,r}$ in Eq.~\eqref{eq:EMcol1} denotes a coefficient appearing in the expansion of $\left[\prod_{j\in\partial A}(\cos(x_j) +\sin(x_j))\right]^r$.
Then the coefficient one-norm, $\|\mathbf{a}_r\|_1$, satisfies
\begin{align}
&\|\mathbf{a}_r\|_1 = \sum_{i}|a_{i,r}| = 
\left|[\prod_{j\in\partial A}(|\cos(x_j)|+|\sin(x_j)|)]^r\right| \notag \\
&\leq
\left[\prod_{j\in\partial A}(|\cos(x_j)|+|\sin(x_j)|)\right]^r \notag \\
&\leq \left[\prod_{j\in\partial A}e^{x_j}\right]^r\leq e^{\eta t},
\end{align}
with $\eta \equiv \sum_{j\in\partial A} \|\hat{H}_j\|$.
Here, we used the triangle inequality in the second line and $|\cos (x_j)|+|\sin (x_j)|\leq e^{x_j}$ for $x_j\geq0$. 
Thus $\gamma\leq |C|^{-1}e^{2\eta t}$, and hence the sampling overhead $\gamma^2$ is at most $|C|^{-2}e^{4\eta t}$.
Since this bound is independent of $r$, taking $r\to\infty$ gives the claimed Hamiltonian-evolution estimate.
\end{proof}
\section{Noise settings, fidelity, and yield in numerical demonstration}
For comparison with conventional purification protocols, we use the same benchmark setting as in Ref.~\cite{fujii2009entanglement}: an initial fidelity of $1-\epsilon=0.9$, single-qubit depolarizing noise with $p_1=0.001$, two-qubit depolarizing noise with $p_2=0.01$, and a readout error rate of $p_\text{mes}=0.03$.
The corresponding noise channels are
$\mathcal{E}_1(\rho)=(1-4p_1/3)\rho+(4p_1/3)\sum_{\hat{P}}\hat{P}\rho\hat{P}/4$,
$\mathcal{E}_2(\rho)=(1-16p_2/15)\rho+(16p_2/15)\sum_{\hat{P}_1,\hat{P}_2}(\hat{P}_1\otimes\hat{P}_2)\rho(\hat{P}_1\otimes\hat{P}_2)/16$,
and
$\mathcal{E}_\text{mes}(\rho)=(1-p_\text{mes})\rho+p_\text{mes}\hat{Z}\rho\hat{Z}$
before each $\hat{X}$ measurement, where the Pauli operators in the sums run over
$\{\hat{I},\hat{X},\hat{Y},\hat{Z}\}$.

To quantify the fidelity in the numerical demonstration, we use the entanglement fidelity
\cite{dur2007entanglement,riera2021entanglement}, defined as
$F_\text{purified} =
\mathrm{Tr}[\hat{P}_\text{Bell}\rho_\text{purified}]
/\mathrm{Tr}[\rho_\text{purified}]$.
This corresponds to taking $\mathcal{T}$ to be the identity map, $\hat{A} = \hat{P}_\text{Bell}$, and
$\hat{O} = \hat{P}_\text{Bell}$ in Eq.~\eqref{eq:genrho}. 
Here, $\rho_\text{purified}$ denotes the state after the $N$-party Hadamard test, and $\hat{P}_\text{Bell} = (\ket{\Phi_\text{Bell}}\bra{\Phi_\text{Bell}})^{\otimes n}$.

The standard yield for conventional purification is defined as
$Y(F_\text{purified},F_\text{noisy})=
\prod_{i=1}^{n_\text{round}(F_\text{purified})}
\frac{p_i(F_{i-1})}{K_i}$,
where $n_\text{round}(F_\text{purified})$ is the number of purification rounds required to achieve $F_\text{purified}$, and $K_i$ is the number of noisy Bell states consumed in the $i$th round.
Here, $p_i(F_{i-1})$ is the success probability of the $i$th purification round using noisy Bell states with fidelity $F_{i-1}$, with $F_0 = F_\text{noisy}$
\cite{dur2007entanglement,fujii2009entanglement,riera2021entanglementPRL,riera2021entanglement}.
This yield quantifies the ratio of the number of purified Bell states with fidelity $F_\text{purified}$ to the number of noisy Bell states with fidelity $F_\text{noisy}$.

Since expectation-value purification protocols, i.e., virtual purification protocols, require a sampling overhead of $[\gamma(n)]^{2}$ to achieve the same statistical accuracy as
conventional purification protocols, we replace the physical success probability with the inverse sampling overhead as an effective success factor.
We therefore define the effective yield of our protocol as
\begin{equation}
Y_\text{virtual}
=
n
\prod_{i=1}^{n_\text{round}(F_\text{purified})}\frac{1}{K_i}
\prod_{\ell=1}^{n}\frac{1}{[\gamma^{(\ell)}]^2}.
\end{equation}
Here, the prefactor $n$ accounts for the number of virtually prepared Bell states, and $[\gamma^{(\ell)}]^2$ denotes the sampling overhead for the $\ell$th virtual step.
More generally, virtual purification can be combined with conventional purification: if a conventional purification protocol is applied before the virtual step, the factor $\prod_i 1/K_i$ is replaced by the corresponding conventional yield $\prod_i p_i(F_{i-1})/K_i$.
Thus, the effective yield of a hybrid protocol is obtained by multiplying the yield of the physical purification step by the inverse sampling overhead of the virtual step.

\section{Additional numerical demonstration with dephasing and amplitude-damping noise}
\begin{figure}
    \centering
    \includegraphics[width=0.9\linewidth]{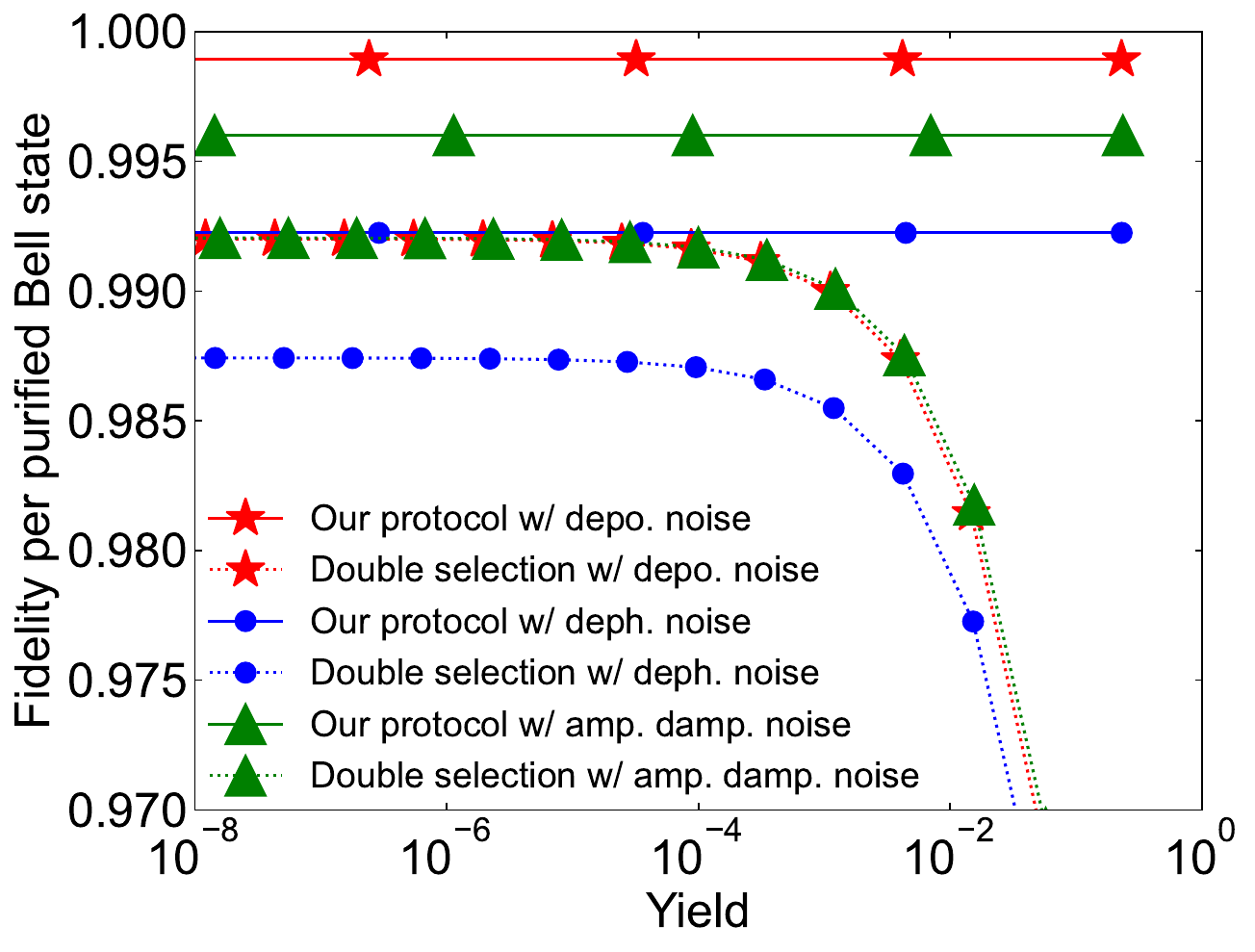}
    \caption{Purified geometric-mean fidelity, i.e., fidelity per purified Bell state, with the corresponding yield for our protocol and the conventional double-selection protocol under depolarizing, dephasing, and amplitude-damping noise.}
   \label{fig:numerics_noise}
\end{figure}

Besides the depolarizing-noise benchmark in the main text, we compare the conventional double-selection protocol with our protocol under two other local noise models in Fig.~\ref{fig:numerics_noise}: dephasing noise applied independently to each qubit, defined by 
$\mathcal{E}_\text{deph}\otimes \mathcal{E}_\text{deph}$ with $\mathcal{E}_\text{deph}(\rho) = (1-p_2')\rho + p_2'\hat{Z}\rho\hat{Z}$, and 
amplitude-damping noise applied independently to each qubit, defined by $\mathcal{E}_\text{amp}\otimes \mathcal{E}_\text{amp}$ with $\mathcal{E}_\text{amp}(\rho) = \sum_{i=0,1}\hat{K}_i\rho\hat{K}_i^\dagger$, where
\begin{equation}
\hat{K}_0 = \begin{pmatrix} 1 & 0 \\ 0 & \sqrt{1-p_2'} \end{pmatrix}, \quad
\hat{K}_1 = \begin{pmatrix} 0 & \sqrt{p_2'} \\ 0 & 0 \end{pmatrix}.
\end{equation}
To compare the noise models at a comparable error strength, we set $1-p_2 = (1-p_2')^2$ with $p_2 = 0.01$.
Figure~\ref{fig:numerics_noise} shows the fidelity and yield under these noise models, indicating that our protocol achieves higher fidelity than the conventional double-selection protocol in all cases.
\section{Probabilistic error cancellation for Bell-state preparation using only LOCC}
Probabilistic error cancellation (PEC) is a standard error-mitigation method \cite{temme2017error,endo2018practical,endo2020hybrid} that implements an inverse noise map by classical post-processing at the cost of sampling overhead.
If $\mathcal{E}^{-1}=\sum_i\eta_i\mathcal{B}_i$ with $\eta_i\in\mathbb{R}$, PEC samples $\mathcal{B}_i$ with probability $p_i=|\eta_i|/\gamma$, where $\gamma=\sum_i|\eta_i|$, and multiplies the outcome by $\gamma\mathrm{sgn}(\eta_i)$.
The required number of shots scales as $\gamma^2$ \cite{endo2018practical}.

For the Werner-state noise model used in the main text, we use the local-Pauli inverse map $\mathcal{E}^{-1}(\rho)=\sum_{i=0,1,2,3}\eta_i(\hat{I}\otimes\hat{P}_i)\rho(\hat{I}\otimes\hat{P}_i)$, where $\eta_0=(3-\epsilon)/(3-4\epsilon)$, $\eta_1=\eta_2=\eta_3=-\epsilon/(3-4\epsilon)$, and $(\hat{P}_0,\hat{P}_1,\hat{P}_2,\hat{P}_3)=(\hat{I},\hat{X},\hat{Y},\hat{Z})$.
The corresponding sampling-cost factor is $\gamma=\sum_{i=0,1,2,3}|\eta_i|=1+6\epsilon/(3-4\epsilon)$.
Although Ref.~\cite{yuan2023virtual} gives an inverse map attaining the optimal PEC cost, we use this local-Pauli inverse map as a practical LOCC benchmark.

The protocol is as follows.
\begin{enumerate}
[  labelindent=0pt,
  leftmargin=*,
  labelsep=.5em,
  topsep=0pt,
  partopsep=0pt,
  parsep=0pt,
  itemsep=0.05cm]
\item Share a noisy Bell state, twirl it into a Werner state, and estimate $\epsilon$ to determine $\eta_i$.
\item Sample a Pauli index $i$ with probability $p_i=|\eta_i|/\gamma$ and communicate it between the parties.
\item Apply the local Pauli operation $\hat{I}\otimes\hat{P}_i$ to the shared noisy Bell state.
\item Run the target circuit and multiply the outcome by $\gamma\mathrm{sgn}(\eta_i)$.
\item Average the outcomes to obtain the desired expectation value.
\end{enumerate}

\clearpage
\onecolumngrid
\begin{center}
	\Large
	\textbf{Supplementary Materials for: $N$-party Hadamard Test for Distributed Quantum Computation}
\end{center}

\setcounter{section}{0}
\setcounter{equation}{0}
\setcounter{figure}{0}
\setcounter{table}{0}
\setcounter{page}{1}
\renewcommand{\thesection}{S\arabic{section}}
\renewcommand{\theequation}{S\arabic{equation}}
\renewcommand{\thefigure}{S\arabic{figure}}
\renewcommand{\thetable}{S\arabic{table}}
\newcommand{\SMcite}[1]{[S#1]}
\section{Derivation of the estimator in Theorem~2}
Here we consider the case of $2\hat{M}_\mathrm{Re}$.
\begin{align}
\sum_{i,j}\left(\frac{\|\mathbf{a}\|_1^2a_ia_j}{|a_ia_j|}\right) p_{ij}\braket{2\hat{M}_\mathrm{Re}\otimes\hat{O}}_{ij} &=\sum_{i,j}a_{i}a_j\, 2\mathrm{Re}\!\left(c\mathrm{Tr}[\hat{O}\mathcal{T}(\hat{U}_i\rho\hat{U}_j^\dagger)]\right) \notag \\
    &= \sum_{i,j}a_{i}a_j\left(c\mathrm{Tr}[\hat{O}\mathcal{T}(\hat{U}_i\rho\hat{U}_j^\dagger)]+c^*\mathrm{Tr}[\hat{O}\mathcal{T}(\hat{U}_j\rho\hat{U}_i^\dagger)]\right) \notag \\
    &= (c+c^*)\mathrm{Tr}[\hat{O}\mathcal{T}(\hat{A}\rho\hat{A}^\dagger)] \notag \\
    & =C\mathrm{Tr}[\hat{O}\mathcal{T}(\hat{A}\rho \hat{A}^\dagger)],
\end{align}
where we used $p_{ij}=|a_ia_j|/\|\mathbf{a}\|_1^2$, $\hat{A}=\sum_i a_i\hat{U}_i$ with $a_i\in\mathbb{R}$, and $c+c^*=2\mathrm{Re}(c)=C$.
In the third line, we used the Hermiticity of $\hat{O}$ and the Hermiticity-preserving property of $\mathcal{T}$ to write $\mathrm{Tr}[\hat{O}\mathcal{T}(\hat{U}_i\rho\hat{U}_j^\dagger)]^*=\mathrm{Tr}[\hat{O}\mathcal{T}(\hat{U}_j\rho\hat{U}_i^\dagger)]$.
For $2\hat{M}_\mathrm{Im}$, the same calculation replaces the plus sign in the third line by a minus sign and includes a factor $1/i$, yielding $(c-c^*)/i=2\mathrm{Im}(c)=C$.
\section{Reduced sampling overhead with known coherence}
In Theorem~2 in the main text, we need to measure an ancilla even when estimating $\mathrm{Tr}[\hat{O}\mathcal{T}(\hat{U}_i\rho\hat{U}_i^\dagger)]$, a quantity that can be evaluated using only a unitary operation $\hat{U}_i$ on the system without ancilla.
This is because $C$ is not assumed to be known in advance, and the protocol is designed so that the same factor $C$ appears in both the numerator and denominator estimators, allowing it to cancel in the ratio.
However, if $C$ is known in advance, then for $i=j$ we can use the unitary operation instead and reduce the sampling overhead by incorporating $C$ into the sampling probability.
The complete protocol is given below. 

\begin{lemma}
Suppose that a nonzero signed coherence factor $C=2\mathrm{Re}(c)$ or $C=2\mathrm{Im}(c)$ is known before the calculation. Then, the sampling-cost factor can be reduced to $\gamma = \mathrm{Tr}[\hat{A}\rho\hat{A}^\dagger]^{-1}D$ with $D=|C|^{-1}\|\mathbf{a}\|_1^2 -(|C|^{-1}-1)\|\mathbf{a}\|_2^2$.
\end{lemma}
\begin{proof}
We prove the lemma by giving an explicit estimator.
\begin{enumerate}[
  labelindent=0pt,
  leftmargin=*,
  labelsep=.5em,
  topsep=0pt,
  partopsep=0pt,
  parsep=0pt,
  itemsep=0.05cm
]
    \item Sample $i,j$ with probability $p_{ij}$, where $p_{ij}=|C|^{-1}|a_ia_j|/D$ for $i\neq j$ and $p_{ij}=|a_ia_j|/D$ for $i=j$.
    One can verify directly from the definition of $D$ that this distribution is normalized.
    \item For $i\neq j$, run the $N$-party Hadamard test with $\hat{U} = \hat{U}_i =\bigotimes_{k=1}^N \hat{U}_i^{(k)}$, $\hat{V} = \hat{U}_j = \bigotimes_{k=1}^N \hat{U}_j^{(k)}$, and the measurement $2\hat{M}_\mathrm{Re}$ or $2\hat{M}_\mathrm{Im}$ together with $\hat{O}$. 
    For $i=j$, apply the unitary $\hat{U}_i$ to the system, followed by $\mathcal{T}$ and the measurement of $\hat{O}$.
    \item Multiply the obtained expectation value by $D a_ia_j/|a_ia_j| \times \mathrm{sgn}(C)$ for $i\neq j$ and by $D a_ia_j/|a_ia_j|$ for $i=j$, and average over $i,j$.
    This gives an estimator of the numerator of Eq.~\eqref{eq:genrho} as 
    $\sum_{i\neq j} (D a_ia_j/|a_ia_j|)\mathrm{sgn}(C)p_{ij}\braket{2\hat{M}_\mathrm{Re(Im)}\otimes\hat{O}}_{ij}  + \sum_{i=j}(D a_ia_j/|a_ia_j|)p_{ij}\braket{\hat{O}}_{ij}
    =\mathrm{Tr}[\hat{O}\mathcal{T}(\hat{A}\rho \hat{A}^\dagger)]$, whose detailed derivation is similar to that in the previous section.
    The multiplication increases the variance by a factor of $D^2$. 
    \item Estimate the denominator from the same data by discarding the measurement result of $\hat{O}$ for $i\neq j$ and setting $\braket{\hat{O}}_{ii}=1$ for $i=j$.
    This is equivalent to setting $\hat{O} = \hat{I}$ in the above procedure, and therefore gives $\mathrm{Tr}[\mathcal{T}(\hat{A}\rho \hat{A}^\dagger)] =\mathrm{Tr}[\hat{A}\rho \hat{A}^\dagger]$, since $\mathcal{T}$ is trace-preserving.
    No additional quantum processing is required.
    \item Dividing the estimator for $\mathrm{Tr}[\hat{O}\mathcal{T}(\hat{A}\rho \hat{A}^\dagger)]$ by that for $\mathrm{Tr}[\hat{A}\rho \hat{A}^\dagger]$ provides the desired value. 
\end{enumerate}

In the asymptotic large-sample limit, the variance of the ratio estimator $\bar{x}/\bar{y}$, constructed from the sample means $\bar{x}$ and $\bar{y}$, is given by $\mathrm{Var}(\bar{x}/\bar{y}) = \sigma_\mathrm{ratio}^2N_{\mathrm{samp}}^{-1} + \mathcal{O}(N_{\mathrm{samp}}^{-2})$, where $\sigma_\mathrm{ratio}^2 = \mu_y^{-2}[\mathrm{Var}(x) + \mu_x^2\mu_y^{-2}\mathrm{Var}(y) - 2\mu_x\mu_y^{-1}\mathrm{Cov}(x,y)]$, $N_{\mathrm{samp}}$ is the number of samples, $\mu_x$ and $\mu_y$ denote the means of $x$ and $y$, respectively, and $\mathrm{Cov}(x,y)$ is the covariance \SMcite{1}. Thus, compared with a standard quantum observable, achieving the same variance requires increasing the number of samples by a factor of $\sigma_\mathrm{ratio}^2$.
For the protocol described above, we have $\sigma_\mathrm{ratio}^2= \mathcal{O}(\gamma^2)$ up to a constant factor since $\mathrm{Var}(x)$ and $\mathrm{Var}(y)$ scale as $D^2$ due to the multiplication in Step 3, and $\mu_y = \mathrm{Tr}[\hat{A}\rho\hat{A}^\dagger]$.
\end{proof}
\begin{remarks}
\begin{itemize}[
  labelindent=0pt,
  leftmargin=*,
  labelsep=.5em,
  topsep=0pt,
  partopsep=0pt,
  parsep=0pt,
  itemsep=0.05cm
]
\item For $N=2$ with the separable ancilla $\rho_\text{anc} = (\ket{+}\bra{+})^{\otimes 2}$, where $|C|=1/2$, and for unitary $\hat{A}$, $\gamma$ reduces to the ``product extent'' defined in Ref.~\SMcite{2}.
\end{itemize}
\end{remarks}

\section*{References}

\begin{enumerate}[label={[S\arabic*]}, leftmargin=*]

\item A.~W.~van der Vaart,
{\it Asymptotic Statistics},
Cambridge Series in Statistical and Probabilistic Mathematics, Vol.~3,
Cambridge University Press, Cambridge (1998).

\item A.~W.~Harrow and A.~Lowe,
``Optimal Quantum Circuit Cuts with Application to Clustered Hamiltonian Simulation,''
PRX Quantum \textbf{6}, 010316 (2025).

\end{enumerate}

\end{document}